\documentclass[aps,pra,twocolumn,groupedaddress,showpacs,floatfix]{revtex4}
\usepackage{graphicx}
\usepackage{amsmath}

\begin{document}

\title{Observation of conformal symmetry breaking and scale invariance\\in expanding Fermi gases}

\author{E. Elliott$^{1,2}$, J. A. Joseph$^1$, and J. E.Thomas$^1$}

\affiliation{$^{1}$Department of  Physics, North Carolina State University, Raleigh, NC 27695, USA}
\affiliation{$^{2}$Department of Physics, Duke University, Durham, NC 27708, USA}


\date{\today}

\begin{abstract}
We precisely test scale invariance and local thermal equilibrium in the hydrodynamic expansion of a Fermi gas of atoms as a function of interaction strength.   After release from an anisotropic optical trap, we observe that a resonantly interacting gas  obeys scale-invariant hydrodynamics, where the mean square cloud size   $\langle{\mathbf{r}}^2\rangle=\langle x^2+y^2+z^2\rangle$ expands ballistically (like a noninteracting gas) and the energy-averaged bulk viscosity is consistent with zero, $0.005(0.016)\,\hbar\,n$, with $n$ the density. In contrast, the aspect ratios of the cloud exhibit anisotropic ``elliptic" flow with an energy-dependent shear viscosity.   Tuning away from resonance, we observe conformal symmetry breaking, where  $\langle{\mathbf{r}}^2\rangle$  deviates from ballistic flow.
\end{abstract}

\maketitle

The identification and comparison of scale-invariant physical systems, defined as those without an intrinsic length scale, has enabled significant advances connecting diverse fields of physics. Of recent interest are connections between scale-invariant strongly interacting systems and their weakly interacting counterparts. An important example is the Anti de Sitter-Conformal Field Theory correspondence,  which links a broad class of strongly interacting quantum fields to weakly interacting gravitational fields in five dimensions~\cite{NJPReview}. This correspondence has been used to predict a universal lower bound for the ratio of shear viscosity to entropy density~\cite{KovtunViscosity}, connecting quark-gluon-plasmas~\cite{Heinz,Shuryak} to ultra-cold Fermi gases~\cite{SchaeferRatio,CaoViscosity,CaoNJP,NJPReview}. An ultra-cold  Fermi gas is a paradigm for scale-invariant quantum fluids  with the unique trait that a cloud of spin-up and spin-down atoms is magnetically tunable between scale-invariant strongly interacting and noninteracting fluids. The development of non-relativistic conformal field theory~\cite{SonConformalFermi} may expose  a deep geometric correspondence between these two regimes.

Near a collisional (Feshbach) resonance, the s-wave scattering length $a_S$ for interactions between spin-up and spin-down atoms can be tuned to a zero crossing, where $a_S=0$ and the gas is noninteracting. Tuning to resonance, where $a_S$ diverges, the cloud is the most strongly interacting, non-relativistic quantum system known~\cite{SchaferMostStrongly}. A central connection between these two regimes is that in both cases, the  thermal equilibrium pressure $p$ and energy density ${\cal E}$ are related by $p=\frac{2}{3}\, {\cal E}$, which follows from the universal hypothesis~\cite{HoUniversalThermo,ThomasUniversal}. This equation of state for a resonantly interacting Fermi gas has been verified experimentally to high precision~\cite{KuThermo}, but only for a trapped gas. An obvious distinction between the ideal and strongly-interacting regimes was first demonstrated  by observing  the aspect ratio of a Fermi gas after release from an anisotropic trap~\cite{OHaraScience}. The ideal gas was shown to expand ballistically with an isotropic momentum distribution, whereas  the strongly interacting gas was found to expand hydrodynamically and to exhibit anisotropic ``elliptic" flow~\cite{Heinz,OHaraScience}.

In this Letter,  we demonstrate both theoretically and experimentally that  scale-invariance connects the resonantly interacting and ideal noninteracting gas by requiring the mean square cloud size $\langle {\mathbf{r}}^2\rangle=\langle x^2+y^2+z^2\rangle$ to expand identically, in contrast to the aspect ratios.   Tuning the cloud  away from resonance, where the scattering length is finite, we observe  breaking of scale invariance, which is controlled by the conformal symmetry breaking pressure $\Delta p=p-\frac{2}{3}\, {\cal E}$ and a finite bulk viscosity. We show that measurement of $\langle {\mathbf{r}}^2\rangle=\langle x^2+y^2+z^2\rangle$ enables a precision measurement of the bulk viscosity without creating a spherical trap and tests local thermal equilibrium during expansion.
\begin{figure}
\begin{center}\
\includegraphics[width=2.75in]{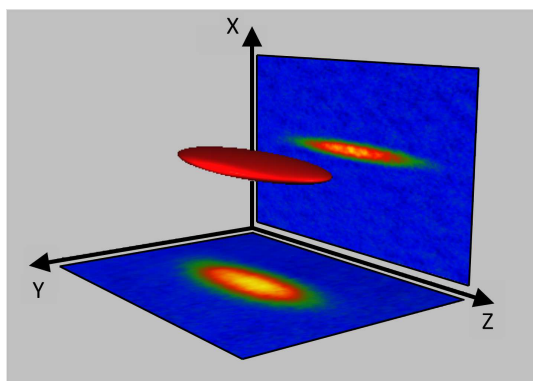}
\end{center}
\caption{Imaging the expanding cloud in three dimensions. Two CCD cameras are used to measure the density profile of the cloud. The cloud is released from an asymmetric optical trap with a 1.0\,:\,2.7\,:\,33 (x:y:z) aspect ratio, enabling observation of elliptic flow in the $x-y$ plane.\label{fig:apparatus}}
\end{figure}

In the experiments, we employ an optically-trapped cloud of $^6$Li atoms in a 50-50 mixture of the two lowest hyperfine states, which is cooled by evaporation~\cite{OHaraScience}. We  determine $\widetilde{E}\equiv\langle {\mathbf{r}}\cdot\nabla U\rangle_0$ from the trapped cloud profile and use it as an interaction-independent initial energy scale~\cite{SupportOnline}. The cloud is released from an anisotropic trap with a 1:2.7:33  aspect ratio. Two independent images,  Fig.~\ref{fig:apparatus}, are obtained using two CCD cameras and two simultaneous, orthogonally-propagating probe beam pulses, which each interact with a different hyperfine state. In this way, the cloud profile is measured as a function of time after release in all three dimensions.

 We relate the acceleration of the mean square cloud radius to the conformal symmetry breaking pressure $\Delta p$ and the bulk viscosity $\zeta_B$, using the hydrodynamic equation for the velocity field ${\mathbf{v}}$ (including pressure and viscous forces) and the continuity equation for the density $n$, which are consistent with energy conservation. {\it Without} assuming a scaling solution, we find that a single-component fluid comprising $N$ atoms of mass $m$ obeys~\cite{SupportOnline},
\begin{eqnarray}
\frac{d^2}{dt^2}\frac{m\langle{\mathbf{r}}^2\rangle}{2}&=&\langle{\mathbf{r}}\cdot\nabla U_{opt}\rangle_0+\frac{3}{N}\int d^3{\mathbf{r}}\,[\Delta p-\Delta p_{\,0}]\nonumber\\
& & -\frac{3}{N}\int d^3{\mathbf{r}}\,\zeta_B\nabla\cdot{\mathbf{v}},
\label{eq:rsq}
\end{eqnarray}
where the subscript ($_0$) denotes the condition at $t=0$, just after the optical trap is extinguished and $\zeta_B$ is the local bulk viscosity.
For brevity, we include only the optical trap potential $U_{opt}$ in Eq.~\ref{eq:rsq}, which need not be harmonic. However, for our precision measurements, as described below, it is important to include also the small potential energy arising from the finite curvature of the bias magnetic field~\cite{SupportOnline}.   As $\langle{\mathbf{r}}^2\rangle$ is a scalar, the contribution of the shear viscosity pressure tensor vanishes, since it is  traceless.

\begin{figure}
\begin{center}\
\includegraphics[width=3in]{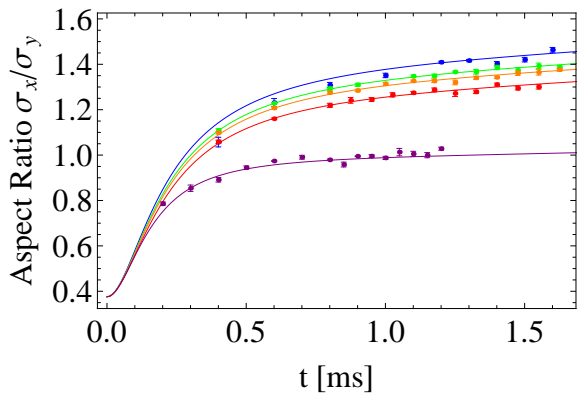}
\end{center}
\caption{Transverse aspect ratio $\sigma_x/\sigma_y$ versus time after release showing elliptic hydrodynamic flow: Top to bottom, resonantly interacting gas at 834 G, $\widetilde{E}=0.66\,E_F$, $\widetilde{E}=0.89\,E_F$, $\widetilde{E}=1.17\,E_F$, $\widetilde{E}=1.46\,E_F$, ballistic (noninteracting) gas at 528 G, $\widetilde{E}=1.78\,E_F$.  Top four solid curves: Hydrodynamic theory with the shear viscosity as the only fit parameter; Lower solid curve: Ballistic theory with no free parameters.  Error bars denote statistical fluctuations in the aspect ratio. \label{fig:aspectratio}}
\end{figure}

\begin{figure}
\begin{center}\
\includegraphics[width=3in]{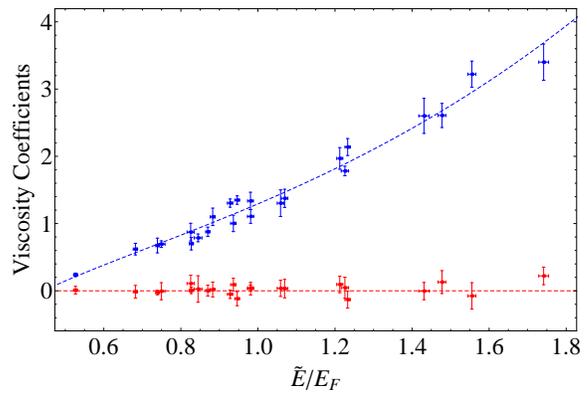}
\end{center}
\caption{Measurement of bulk and shear viscosity for a scale-invariant Fermi gas: Blue (top): Trap-averaged shear viscosity coefficient $\int d^3{\mathbf{r}}\,\eta/(N\hbar)\equiv \bar{\alpha}_S$ versus energy $\widetilde{E}/E_F$. Red (bottom): Trap-averaged bulk viscosity coefficient $\int d^3{\mathbf{r}}\,\zeta_B/(N\hbar)\equiv \bar{\alpha}_B$ versus energy. The weighted average value of $\bar{\alpha}_B=0.005(0.016)$ is consistent with zero. (Dotted curves added to guide the eye.)\label{fig:viscosity}}
\end{figure}

The  aspect ratio $\sigma_x/\sigma_y$ of the cloud is measured at the Feshbach resonance (834 G) as a function of time after release to establish that the flow is hydrodynamic and  to determine the shear viscosity. Fig.~\ref{fig:aspectratio} shows data for  $\widetilde{E}/E_F=0.66,0.89,1.17,1.46$,  where $E_F\equiv \frac{\hbar^2k_{FI}^2}{2m}$ is the measured Fermi energy of an ideal gas at the trap center with $k_{FI}$ the corresponding wavevector~\cite{SupportOnline}. The hydrodynamic expansion data at 834 G is compared to that of a noninteracting gas taken at 528 G where $a_S=0$ and $\widetilde{E}/E_F=1.78$. For the noninteracting gas, which expands ballistically,  the aspect ratio saturates to unity. In contrast, for the resonantly interacting cloud,   $\sigma_x/\sigma_y$ increases to approximately $1.5$ over the time range shown, clearly demonstrating that the cloud expands hydrodynamically.  The shear viscosity increases with increasing energy (see Fig.~\ref{fig:viscosity}), slowing down the rate at which the aspect ratio increases with time.

\begin{figure}
\begin{center}\
\includegraphics[width=3in]{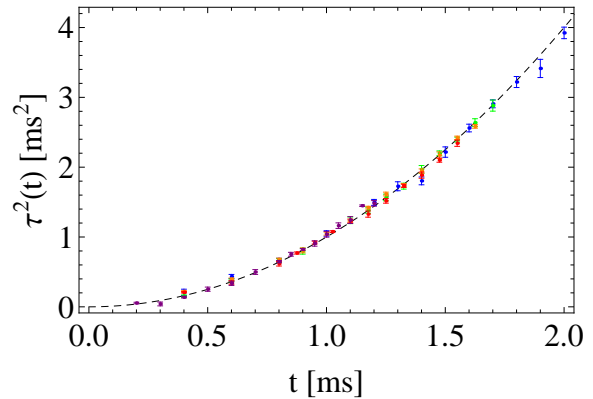}
\end{center}
\caption{Scale invariant expansion of a resonantly interacting Fermi gas. Experimental values of
$\tau^2(t)\equiv m[\langle{\mathbf{r}}^2\rangle-\langle{\mathbf{r}}^2\rangle_0]/\langle{\mathbf{r}}\cdot\nabla U\rangle_0$ versus time $t$ after release, for the same data as in Fig.~\ref{fig:aspectratio} (including noninteracting gas data) collapse onto a single curve, demonstrating universal $t^2$ scaling. Dashed curve $\tau^2(t)=t^2$, as predicted by Eq.~\ref{eq:scaleinv}~\cite{NoteMag}. \label{fig:scaleinv}}
\end{figure}

For a resonantly interacting cloud, important questions are whether $\Delta p=p-\frac{2}{3}\, {\cal E}$ remains zero during expansion and if the expansion is scale-invariant. The bulk viscosity $\zeta_B$  is predicted to vanish in the scale-invariant regime~\cite{SonBulkViscosity,EscobedoBulkViscosity,StringariBulk,DuslingSchaferBulk}, consistent with the bulk viscosity frequency sum rule, which  vanishes when $\Delta p=0$~\cite{RanderiaSum}. If these conditions hold,  Eq.~\ref{eq:rsq} yields
\begin{equation}
\langle{\mathbf{r}}^2\rangle=\langle{\mathbf{r}}^2\rangle_0+\frac{t^2}{m}\,\langle{\mathbf{r}}\cdot\nabla U_{opt}\rangle_0,
\label{eq:scaleinv}
\end{equation}
which corresponds to ballistic expansion of the mean square cloud size (in the same way as a noninteracting gas), even though the individual cloud radii expand hydrodynamically and exhibit elliptic flow, as shown in Fig.~\ref{fig:aspectratio} for the transverse aspect ratio.

Scale invariance of the expanding gas is now directly tested by measuring $\tau^2(t)\equiv m[\langle{\mathbf{r}}^2\rangle-\langle{\mathbf{r}}^2\rangle_0]/\langle{\mathbf{r}}\cdot\nabla U_{opt}\rangle_0$, which should obey $\tau^2(t)=t^2$ for a scale-invariant system, according to Eq.~\ref{eq:scaleinv}~\cite{SupportOnline}. Fig.~\ref{fig:scaleinv} shows the experimental values of $\tau{\,^2}(t)$ versus $t$ for the same data as used in Fig.~\ref{fig:aspectratio}.  In contrast to the aspect ratio versus time data of Fig.~\ref{fig:aspectratio}, which vary substantially with energy due to the shear viscosity, the experimental values of $\tau{\,^2}(t)$ for all energies fall on a single $t^2$ curve, consistent with $\Delta p=0$ and scale invariant expansion.

\begin{figure}
\begin{center}\
\includegraphics[width=3in]{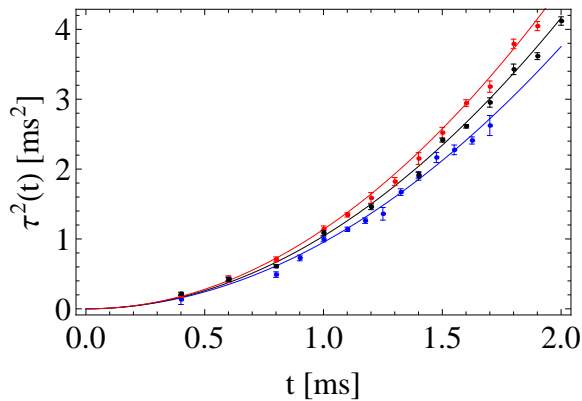}
\end{center}
\caption{Conformal symmetry breaking  in the expansion for a Fermi gas near a Feshbach resonance. The data are the experimental values of $\tau^2(t)\equiv m[\langle{\mathbf{r}}^2\rangle-\langle{\mathbf{r}}^2\rangle_0]/\langle{\mathbf{r}}\cdot\nabla U\rangle_0$ for $\widetilde{E}/E_F\simeq 1.0$, versus time $t$ after release. Solid curves  are the predictions using Eq.~\ref{eq:rsq} with $\zeta_B=0$, where the pressure change $\Delta p$  is approximated using the second virial coefficient without any free parameters~\cite{SupportOnline}.  Top: $1/(k_{FI}a)=-0.59$; Center: $1/(k_{FI}a)=0$; Bottom: $1/(k_{FI}a)=+0.61 $. \label{fig:breakscaleinv}}
\end{figure}

We investigate the breaking of scale invariance and  local  thermodynamic equilibrium  of the expanding gas at finite scattering length by tuning the bias magnetic field above and below the Feshbach resonance. Fig.~\ref{fig:breakscaleinv} shows $\tau^2(t)$ data for $\widetilde{E}\simeq 1.0\,E_F$.  Compared to the resonant case, we see qualitatively that the cloud expands more rapidly when the scattering length is negative and more slowly when the scattering length is positive. This behavior is a signature of the $[\Delta p-\Delta p_{\,0}]$ term in Eq.~\ref{eq:rsq}, where $|\Delta p  (t)|\leq |\Delta p (0)|$ for any time $t$ after release and $\Delta p$ has the same sign as the scattering length.

To estimate $\Delta p-\Delta p_{\,0}$ in Eq.~\ref{eq:rsq}, we employ for simplicity a high-temperature, second virial coefficient approximation~\cite{HoMuellerHighTemp}. We retain only the translational degrees of freedom and ignore the contribution from changes in the molecular population, which require three-body collisions that occur with low probability during the expansion time scale. In $\Delta p$,  the translational temperature  is evaluated using an adiabatic approximation,  so that $\Delta p(t)$ is then a known function of time and is odd in $1/a_S$~\cite{SupportOnline}. We find that estimating $\Delta p$ in this way  yields satisfactory agreement with the data of Fig.~\ref{fig:breakscaleinv}, even for relatively low energies $\widetilde{E}/E_F\simeq 1$. The reasonable fits suggests that two-body interactions are dominant and that the translational degrees of freedom are near local thermal equilibrium in the expanding cloud. The expansion at finite $1/a_S$ is energy dependent, since $\Delta p$ rapidly approaches zero as the energy is increased~\cite{SupportOnline}.

 We present a new precision measurement of the shear viscosity at resonance, which serves as a reference for the bulk viscosity measurement described below. This is accomplished by measuring the {\it transverse} aspect ratio as a function of time after release, Fig.~\ref{fig:aspectratio}. The shear viscosity pressure tensor slows the flow in the  rapidly expanding, initially narrow, $x$-direction and increases the speed in the more slowly expanding $y$-direction. As the initial transverse aspect ratio is 1:2.7 for our trap, elliptic flow is observed for relatively short expansion times with high signal to background ratio, enabling high sensitivity to the shear viscosity, even at the lowest energies, which were not accessible in our previous expansion measurements~\cite{CaoViscosity,CaoNJP}.   We fit the data of  Fig.~\ref{fig:aspectratio}  for a  resonantly interacting gas at 834 G, using a general, energy-conserving hydrodynamic model~\cite{CaoViscosity,CaoNJP}, valid in the scale-invariant regime where $\Delta p=0$.  At resonance, the shear viscosity $\eta$ takes the form $\eta=\alpha_S\,\hbar\,n$, where $n$ is the total density of  atoms  and $\alpha_S$ is a dimensionless function of the local reduced temperature. The trap-averaged shear viscosity coefficient $\bar{\alpha}_S\equiv\int d^3{\mathbf{r}}\,\eta/(\hbar N)$ is used as the only free parameter, initially neglecting the bulk viscosity, which is expected to be much smaller. For the shear viscosity in the scale-invariant regime, $\bar{\alpha}_S$ is an adiabatic invariant, which is therefore temporally constant in the adiabatic approximation~\cite{CaoViscosity,CaoNJP}. The fits to the aspect ratio obtained in this way are shown in Fig.~\ref{fig:aspectratio} as solid lines and yield the data shown in Fig.~\ref{fig:viscosity}.

The bulk viscosity is measured with high sensitivity from the expansion of $\langle{\mathbf{r}}^2\rangle$, Eq.~\ref{eq:rsq}, which is independent of the shear viscosity. The divergence of the velocity field ${\mathbf{v}}$ is easily determined from the fits to the aspect ratios using a scaling approximation, which is adequate for the small bulk viscosity term. Fig.~\ref{fig:scaleinv}  shows that both $\Delta p$ and $\zeta_B$ must be nearly zero. To estimate the bulk viscosity at resonance, we assume that $\Delta p=p-\frac{2}{3}\,{\cal E}=0$ for the expanding, resonantly interacting  gas, so that  the bulk viscosity term in Eq.~\ref{eq:rsq} produces the only deviation from scale-invariance in the evolution of $\langle{\mathbf{r}}^2\rangle$.

Analogous to the shear viscosity, we take the bulk viscosity to be of the form $\zeta_B=\alpha_B\,\hbar\,n$, where $\alpha_B$ is dimensionless, and consider first a large finite scattering length.  Since the bulk viscosity must be positive, the leading contribution in powers of the inverse scattering length $a_S$ must be of the form $\zeta_B=f_B(\theta)\,\hbar\,n/(k_F\,a_S)^2$, where $k_F=(3\pi^2n)^{1/3}$ is the local Fermi wavevector.  Here $f_B(\theta)$ is a dimensionless function of the reduced temperature $\theta$, which is an adiabatic invariant, and hence time-independent in the adiabatic approximation that we use for the small bulk viscosity contribution. As the cloud expands, the density decreases as  $n\propto 1/\Gamma$ in the scaling approximation, where the fits to the aspect ratio data in all three dimensions accurately determine the volume scale factor $\Gamma(t)$~\cite{SupportOnline}. Since  $1/k_F^2\propto \Gamma^{2/3}$, the trap-averaged bulk viscosity coefficient, $\bar{\alpha}_B\equiv\int d^3{\mathbf{r}}\,\zeta_B/(\hbar N)$ is time-dependent and scales as
\begin{equation}
\bar{\alpha}_B(t)=\bar{\alpha}_B(0)\,\Gamma^{2/3}(t).
\label{eq:bulk}
\end{equation}
With the scaling approximation $\nabla\cdot{\mathbf{v}}=\dot{\Gamma}/\Gamma$, the bulk viscosity term then takes the simple form $-3\,\hbar\,\bar{\alpha}_B(0)\,\dot{\Gamma}/\Gamma^{1/3}$. We determine  $\bar{\alpha}_B(0)$ with high precision, by using a least-squares fit of Eq.~\ref{eq:rsq} to the measured $\langle{\mathbf{r}}^2\rangle$ data~\cite{SupportOnline}. In contrast to the shear viscosity coefficient, which increases with increasing energy, Fig.~\ref{fig:viscosity} shows that the bulk viscosity coefficient at resonance remains nearly zero over the entire energy range. We find that the weighted-average $\bar{\alpha}_B(0)=0.005(0.016)$, which is consistent with zero, as predicted for a scale-invariant cloud\cite{SonBulkViscosity,EscobedoBulkViscosity,StringariBulk,DuslingSchaferBulk}. This null result for the bulk viscosity is two orders of magnitude more stringent than that obtained from our previous consistency argument~\cite{CaoNJP}, where only one relatively high energy ($E/E_F\simeq 3.3$) was studied with low sensitivity, by measuring the expansion of the aspect ratio.

We also estimate the bulk viscosity for finite $1/a_S$. From  Fig.~\ref{fig:breakscaleinv}, we see that $\Delta p$, which is an odd function of $1/a_S$, adequately accounts for most of the deviation from scale invariant ballistic expansion. A nonzero bulk viscosity would shift both finite $a_S$ curves  downward. In a more detailed analysis~\cite{SupportOnline}, we fit the data for finite $1/a_S$ by scaling the predicted high temperature $\Delta p$ by a factor $\lambda_p$. We also scale a recent prediction for the high temperature bulk viscosity~\cite{DuslingSchaferBulk} by $\lambda_B$. Our best fits gives  $\lambda_p=1.06(0.21)$ and $\lambda_B=0.21(0.60)$. The bulk measurement is consistent with zero, but places a constraint on the maximum value that is within the range of the prediction~~\cite{DuslingSchaferBulk}.

This research is supported by the Physics Division of the National Science Foundation (Quantum transport in strongly interacting Fermi gases PHY-1067873) and by the Division of Materials Science and Engineering,  the Office of Basic Energy Sciences, Office of Science, U.S. Department of Energy (Thermodynamics in strongly correlated Fermi gases DE-SC0008646). Additional support is provided by the Physics Divisions of the Army Research Office (Strongly interacting Fermi gases in reduced dimensions W911NF-11-1-0420) and the Air Force Office of Scientific Research (Non-equilibrium Fermi gases FA9550-13-1-0041).  The authors are pleased to acknowledge K. Dusling and T. Sch\"{a}fer, North Carolina State University, for stimulating conversations.


\appendix
\section{Supplemental Material}
\label{supplement}

\subsection{Experimental Methods}

In the experiments, we employ an optically-trapped cloud of $^6$Li atoms in a 50-50 mixture of the two lowest hyperfine states, which is tuned to a broad collisional (Feshbach) resonance in a bias magnetic field of 834 G, and cooled by evaporation. The initial energy per particle $\widetilde{E}$ is measured from the trapped cloud profile, as discussed below.   A focused CO$_2$ laser beam forms the cigar-shaped  optical trap with a  transverse aspect ratio of 1:2.7, which enables an observation of transverse elliptic flow on a short time scale and a precise measurement of the static shear viscosity even at low temperature, where the shear viscosity is small. The transverse aspect ratio  is controlled by using two sets of cylindrical ZnSe lenses. One set is placed just after the acousto-optic modulator that controls the laser intensity. A second set is placed just before an expansion telescope. The telescope increases the trapping beam radii before focusing into a high vacuum chamber, where the optical trap is loaded from a standard magneto-optical trap. The first set of cylindrical lenses adjusts the transverse aspect ratio of focused beam, while the second set matches of the beam curvatures to achieve a common focal plane.

To observe the expansion dynamics, the cloud is released from the trap and the cloud profile is measured as a function of time after release in all three dimensions using two identical CCD cameras, which simultaneously image different spin states to avoid cross-saturation. The magnifications of the imaging systems are measured by translating the trap focus. The measured magnifications  yield average axial dimensions $\sigma_z$ that are consistent within 1\%. To obtain the most precise measurements of the cloud profile, we adjust the effective magnification of one camera so that the average axial dimensions precisely agree. In this way, the cloud radii $\sigma_i$ in all three dimensions are consistently measured, to determine the aspect ratios $\sigma_x/\sigma_y$, $\sigma_x/\sigma_z$, and $\sigma_y/\sigma_z$, as well as the mean square cloud radius, $\langle {\mathbf{r}}^2\rangle$. Two-dimensional density distributions are fit to the cloud profiles to extract the cloud radii. For fast data handling,  gaussian profiles are assumed for most of the data and a zero-temperature Thomas-Fermi profile is assumed for the lowest energies. Both types of fit profiles are compared to full finite-temperature Thomas-Fermi profiles to estimate  multiplicative corrections to the cloud radii, which are needed to correct for the small error arising from the form of the fit functions.

We  derive an exact, model-independent evolution equation for $\langle {\mathbf{r}}^2\rangle$ based on hydrodynamics and energy conservation in \S~\ref{sec:rsq}. This enables precise characterization of scale-invariance and local thermodynamic equilibrium in an expanding cloud. The primary result, Eq.~\ref{eq:3.1}, is independent of the shear viscosity and includes the corrections to the flow arising from the bulk viscosity and the deviation $\Delta p\equiv p-\frac{2}{3}\,{\cal E}$ of the pressure from the scale invariant equation of state, $p=\frac{2}{3}\,{\cal E}$. We also include the potential energy arising from the finite bias magnetic field curvature, as required for our precision measurements.

The pressure change $\Delta p$ is determined for the high temperature limit in \S~\ref{sec:highT}. Then the method of estimating the bulk viscosity is described in \S~\ref{sec:bulk}. Finally,  in \S~\ref{sec:parameters}, we discuss the method of fitting the mean square size $\langle {\mathbf{r}}^2\rangle$ data, when the bias field is tuned both to resonance and off-resonance in the large scattering length regime.
\subsection{Expansion of the Mean-Square Cloud Radius}
\label{sec:rsq}

We employ a hydrodynamic description for a single component fluid~\cite{CaoViscosity,CaoNJP}, where the velocity field $\mathbf{v}(\mathbf{x},t)$ is determined by the scalar pressure and the viscosity pressure tensor,
\begin{eqnarray}
n\,m\left(\partial_t +\mathbf{v}\cdot\nabla\right)v_i&=&-\partial_i p + \sum_j \partial_j (\eta\,\sigma_{ij}+\zeta_B\,\sigma^{'}\delta_{ij})\nonumber \\
& &-n\,\partial_i U_{total}.
\label{eq:force}
\end{eqnarray}
Here $p$ is the scalar pressure and $m$ is the atom mass.  $U_{total}$ is the total trapping potential energy arising from the optical trap and the bias magnetic field curvature.  The second term on the right describes the friction forces
arising from both shear $\eta$ and bulk $\zeta_B$ viscosities, where $\sigma_{ij}=\partial
v_i/\partial x_j+\partial v_j/\partial
x_i-2\delta_{ij}\nabla\cdot\mathbf{v}/3$  and $\sigma^{'}\equiv\nabla\cdot\mathbf{v}$. Current conservation for the density $n(\mathbf{x},t)$ requires
\begin{equation}
\frac{\partial n}{\partial t}+\nabla\cdot(n{\mathbf{v}})=0.
\label{eq:ncons}
\end{equation}
Finally,  consistent with Eq.~\ref{eq:force} and Eq.~\ref{eq:ncons}, conservation of total energy  is described by
\begin{equation}
\frac{d}{d t}\int d^3{\mathbf{r}}\left(n\frac{1}{2}m{\mathbf{v}}^2+{\cal E}+n\,U_{total}\right)=0.
\label{eq:energycons}
\end{equation}
Here,  the first term is the kinetic energy arising from the velocity field and ${\cal E}$ is the internal energy density of the gas. As shown below, Eq.~\ref{eq:energycons} will play an important role in determining a general evolution equation for the volume integral of the pressure in both the scale-invariant regime and away from scale-invariance.

To explore scale invariance for an expanding cloud without creating a spherical trap, we measure  the mean-square cloud radius, $\langle{\mathbf{r}}^2\rangle$, which is a scalar quantity. In this section, we derive generally the equation of motion for $\langle{\mathbf{r}}^2\rangle$, with no simplifying assumptions, except that of a single component hydrodynamically expanding fluid. This approach is appropriate in the normal fluid regime above the superfluid transition temperature as well as in the superfluid regime when the normal and superfluid components move together~\cite{StringariBulk}. We show that in the scale-invariant regime at a Feshbach resonance, where $p-\frac{2}{3}\,{\cal E}$ and $\zeta_B$ should be $0$, conservation of total energy leads to {\it ballistic} expansion of $\langle{\mathbf{r}}^2\rangle$ for a hydrodynamic gas. Away from resonance, the departure from scale-invariance is determined by the change in the equation of state, characterized by the conformal symmetry breaking pressure $\Delta p\equiv p-\frac{2}{3}{\cal E}$ and a finite bulk viscosity $\zeta_B$.

We begin by noting that for each direction $i=x,y,z$, the mean square size $\langle x_i^2\rangle\equiv\frac{1}{N}\int d^3{\mathbf{r}}\,n({\mathbf{r}},t)\,x_i^2$ obeys
\begin{eqnarray}
\frac{d\langle x_i^2\rangle}{d t}&=&\frac{1}{N}\int d^3{\mathbf{r}}\,\frac{\partial n}{\partial t}x_i^2=\frac{1}{N}\int d^3{\mathbf{r}}\,[-\nabla\cdot(n{\mathbf{v}})]x_i^2\nonumber\\
&=&\frac{1}{N}\int d^3{\mathbf{r}}\,n\,{\mathbf{v}}\cdot\nabla x_i^2=2\langle x_i\,v_i\rangle,
\label{eq:xsq}
\end{eqnarray}
where $N$ is the total number of atoms. We have used integration by parts  and $n=0$ for $x_i\rightarrow\pm\infty$ to obtain the second line. Similarly,
\begin{eqnarray}
\frac{d\langle x_iv_i\rangle}{d t}&=&\frac{1}{N}\int d^3{\mathbf{r}}\,n\,x_i\frac{\partial v_i}{\partial t}+\frac{1}{N}\int d^3{\mathbf{r}}\,\frac{\partial n}{\partial t}\,x_i v_i\nonumber\\
&=&\frac{1}{N}\int d^3{\mathbf{r}}\,n\,x_i\frac{\partial v_i}{\partial t}+\frac{1}{N}\int d^3{\mathbf{r}}\,n\,{\mathbf{v}}\cdot\nabla (x_iv_i)\nonumber\\
&=&\langle x_i(\partial_t+{\mathbf{v}}\cdot\nabla)v_i\rangle+\langle v_i^2\rangle.
\label{eq:xv}
\end{eqnarray}
Combining Eq.~\ref{eq:xsq} and Eq.~\ref{eq:xv}, we obtain,
\begin{equation}
\frac{d^2}{d t^2}\frac{\langle x_i^2\rangle}{2}=\langle x_i(\partial_t+{\mathbf{v}}\cdot\nabla)v_i\rangle+\langle v_i^2\rangle.
\label{eq:xsqddot1}
\end{equation}
To proceed, we use Eq.~\ref{eq:force}, which yields
\begin{eqnarray*}
\int d^3{\mathbf{r}}\,n\,x_i(\partial_t+{\mathbf{v}}\cdot\nabla)v_i&=&\nonumber\\
& &\hspace*{-0.5in}\frac{1}{m}\int d^3{\mathbf{r}}\,x_i(-\partial_i p-n\,\partial_i U_{total})\nonumber\\
& &\hspace*{-0.75in}+\frac{1}{m}\sum_j\int d^3{\mathbf{r}}\,x_i\partial_j (\eta\,\sigma_{ij}+\zeta_B\,\sigma^{'}\delta_{ij})
\end{eqnarray*}
Integrating by parts on the right hand side, assuming that the surface terms vanish, we obtain
\begin{eqnarray}
\langle x_i(\partial_t+{\mathbf{v}}\cdot\nabla)v_i\rangle&=&\frac{1}{Nm}\int d^3{\mathbf{r}}\,p-\frac{1}{m}\langle x_i\partial_iU_{total}\rangle \nonumber \\
& &\hspace*{-0.25in}-\frac{1}{Nm}\int d^3{\mathbf{r}}\,(\eta\,\sigma_{ii}+\zeta_B\,\sigma')
\end{eqnarray}
with $\sigma'\equiv\nabla\cdot{\mathbf{v}}$.
Defining the viscosity coefficients $\alpha_S$ and $\alpha_B$ by $\eta\equiv\alpha_S\,\hbar\,n$ and $\zeta_B\equiv\alpha_B\,\hbar\,n$, respectively, we can write,
\begin{eqnarray}
\langle x_i(\partial_t+{\mathbf{v}}\cdot\nabla)v_i\rangle&=&\frac{1}{Nm}\int d^3{\mathbf{r}}\,p-\frac{1}{m}\langle x_i\partial_iU_{total}\rangle \nonumber \\
& &-\frac{\hbar}{m}\langle\alpha_S\,\sigma_{ii}+\alpha_B\,\sigma'\rangle,
\label{eq:2.4a}
\end{eqnarray}
where
\begin{equation}
\langle\alpha_S\,\sigma_{ii}+\alpha_B\,\sigma'\rangle\equiv\frac{1}{N}\int d^3{\mathbf{r}}\,n\,(\alpha_S\,\sigma_{ii}+\alpha_B\,\sigma').
\end{equation}
Using Eq.~\ref{eq:2.4a} in Eq.~\ref{eq:xsqddot1}, we then obtain for one direction $x_i$,
\begin{eqnarray}
\frac{d^2}{dt^2}\frac{\langle x_i^2\rangle}{2}&=&\frac{1}{Nm}\int d^3{\mathbf{r}}\,p+\langle v_i^2\rangle-\frac{1}{m}\langle x_i\partial_iU_{total}\rangle \nonumber \\
& &-\frac{\hbar}{m}\langle\alpha_S\,\sigma_{ii}+\alpha_B\,\sigma'\rangle.
\label{eq:xsqddot2}
\end{eqnarray}
Eq.~\ref{eq:xsqddot2} determines the evolution of the mean square cloud radii along each axis, $\langle x_i^2\rangle$, which depends on the conservative forces arising from the scalar pressure and the trap potential, as well as the viscous forces arising from the shear and bulk viscosities.

Summing Eq.~\ref{eq:xsqddot2} over all three directions, the shear viscosity term vanishes, since $\sigma_{ij}$ is traceless, yielding
\begin{eqnarray}
\frac{d^2}{d t^2}\frac{\langle {\mathbf{r}}^2\rangle}{2}&=&\frac{3}{Nm}\int d^3{\mathbf{r}}\,p+\langle {\mathbf{v}}^2\rangle-\frac{1}{m}\langle {\mathbf{r}}\cdot\nabla U_{total}\rangle \nonumber \\
& &-\frac{3\hbar}{m}\,\langle\alpha_B\,\nabla\cdot{\mathbf{v}}\rangle.
\label{eq:1.1}
\end{eqnarray}
At $t=0^-$, {\it before} release from the trap, ${\mathbf{v}}=0$, Eq.~\ref{eq:1.1} shows that the volume integral of the pressure is
\begin{equation}
\frac{3}{N}\int d^3{\mathbf{r}}\,p_{\,0}=\langle {\mathbf{r}}\cdot\nabla U_{total}\rangle_0,
\label{eq:1.2}
\end{equation}
where the subscript $(\,)_0$ denotes the initial condition.  Here $U_{total}=U_{opt}+U_{mag}$ is the total trapping potential, comprising an optical component from the laser trap and a magnetic component arising from the curvature of the bias magnetic field used in the experiments, as described further below.

It will be convenient to rewrite Eq.~\ref{eq:1.1} in terms of $\Delta p\equiv p-\frac{2}{3}\,{\cal E}$ using
\begin{equation}
p= \frac{2}{3}\,{\cal E}+\Delta p,
\label{eq:1.4}
\end{equation}
where the first term defines the equation of state for the pressure in the scale-invariant regime, and the second term is the conformal symmetry breaking pressure change. Then,
\begin{eqnarray}
\frac{d^2}{dt^2}\frac{\langle {\mathbf{r}}^2\rangle}{2}&=&\frac{2}{Nm}\int d^3{\mathbf{r}}\,{\cal E}+\langle {\mathbf{v}}^2\rangle+\frac{3}{Nm}\int d^3{\mathbf{r}}\Delta p\nonumber\\
& &\hspace*{-0.125in}-\frac{1}{m}\langle {\mathbf{r}}\cdot\nabla U_{total}\rangle-\frac{3\hbar}{m}\,\langle\alpha_B\,\nabla\cdot{\mathbf{v}}\rangle.
\label{eq:1.5}
\end{eqnarray}

Just after release from the trap, $t\geq 0^+$, from the optical trap, the trapping potential changes abruptly,  $U_{total}\rightarrow U_{mag}$. To determine the evolution of $\langle {\mathbf{r}}^2\rangle$ after release, we use total energy conservation to eliminate $\langle {\mathbf{v}}^2\rangle$ from Eq.~\ref{eq:1.5}. From Eq.~\ref{eq:energycons}, the final total energy per particle is equal to the initial total energy,
\begin{eqnarray}
\frac{1}{N}\int d^3{\mathbf{r}}\,{\cal E}+\frac{m}{2}\langle{\mathbf{v}}^2\rangle+\langle U_{mag}\rangle &=&\nonumber\\
& &\hspace*{-1.25in}\frac{1}{N}\int d^3{\mathbf{r}}\,{\cal E}_0+\langle U_{mag}\rangle_0.
\end{eqnarray}

To determine the initial internal energy, $\frac{1}{N}\int d^3{\mathbf{r}}\,{\cal E}_0$, we use ${\cal E}_0=\frac{3}{2}p_{\,0}-\frac{3}{2}\Delta p_{\,0}$. With Eq.~\ref{eq:1.2}, this yields
\begin{equation}
\frac{1}{N}\int d^3{\mathbf{r}}\,{\cal E}_0=\frac{1}{2}\langle {\mathbf{r}}\cdot\nabla U_{total}\rangle_0-\frac{3}{2N}\int d^3{\mathbf{r}}\Delta p_{\,0}.
\label{eq:2.2}
\end{equation}

We have $\langle {\mathbf{r}}\cdot\nabla U_{total}\rangle_0=\langle {\mathbf{r}}\cdot\nabla U_{opt}\rangle_0+\langle {\mathbf{r}}\cdot\nabla U_{mag}\rangle_0$ in  Eq.~\ref{eq:2.2}. Multiplying Eq.~\ref{eq:2.2} by $2/m$  determines the first two terms in Eq.~\ref{eq:1.5}. Then, with $\langle {\mathbf{r}}\cdot\nabla U_{total}\rangle\rightarrow\langle {\mathbf{r}}\cdot\nabla U_{mag}\rangle$ for $t\geq 0^+$ in Eq.~\ref{eq:1.5}, we obtain finally our central result for studying scale invariance,
\begin{eqnarray}
\frac{d^2}{dt^2}\frac{m\langle {\mathbf{r}}^2\rangle}{2}&=&\langle {\mathbf{r}}\cdot\nabla U_{opt}\rangle_0
+\frac{3}{N}\int d^3{\mathbf{r}}\,[\Delta p-\Delta p_{\,0}] \nonumber \\
& &-3\,\hbar\,\langle\alpha_B\,\nabla\cdot{\mathbf{v}}\rangle+\Delta U_{mag},
\label{eq:3.1}
\end{eqnarray}
where we define the conformal symmetry breaking pressure
\begin{equation}
\Delta p\equiv p-\frac{2}{3}\,{\cal E},
\label{eq:Deltap}
\end{equation}
which describes the departure of the pressure from the scale-invariant regime. We also define
\begin{eqnarray}
\Delta U_{mag}&\equiv& 2\langle U_{mag}\rangle_0+\langle {\mathbf{r}}\cdot\nabla U_{mag}\rangle_0\nonumber\\
& &-2\langle U_{mag}\rangle-\langle {\mathbf{r}}\cdot\nabla U_{mag}\rangle,
\label{eq:DeltaU}
\end{eqnarray}
which corrects for the small potential energy arising from the bias magnetic field curvature. As the bias coils are oriented along the $x$ direction, the effective potential is repulsive along the $x$ axis and twice the magnitude of the attractive potential along $y$ and $z$,
\begin{equation}
U_{mag}({\mathbf{r}})=\frac{1}{2}\,m\,\omega_{mag}^2\left(y^2+z^2-2x^2\right),
\label{eq:Mag}
\end{equation}
where $\omega_{mag}=2\pi\times 21.5(0.25)$ Hz at 834 G is measured  from the oscillation frequency of the cloud in the $y-z$ plane. Note that $\omega_{mag}^2[B]$ is proportional to $B$.

The first three terms in Eq.~\ref{eq:3.1} reproduce the Eq.~1 of the main text, where the small magnetic contribution was neglected for brevity. The potential energy arising from the magnetic field curvature depends on the mean square cloud radii, $\langle x_i^2\rangle$, which are determined as a function of time by fitting the aspect ratio data using a scaling approximation for the density profile.
The cloud radii for $t=0^+$ are dominated by $\langle z^2\rangle_0$, the longest direction of the cigar-shaped cloud in the trap. This is consistently measured both by in-situ imaging and by measurements after expansion, using the calculated expansion factor, which is close to unity.

 $\langle {\mathbf{r}}\cdot\nabla U_{opt}\rangle_0$ is determined by  the trap parameters and measurements of the cloud radii,  \S~\ref{sec:xdotgradU}. We  determine the harmonic oscillation frequencies, $\omega_i$ by parametric resonance methods. We subtract off the contribution from the magnetic potential and extrapolate to the harmonic values of the optical frequencies by correcting for trap anharmonicity. We obtain $\omega_x=2\pi\times 2210(4)$ Hz, $\omega_y=2\pi\times 830(2)$, and $\omega_{z\,opt}=2\pi\times 60.7 (0.1)$. The corresponding trap depth is $U_0=60.3(0.2)\,\mu$K. The Fermi energy of an ideal gas at the trap center is $E_F=(3N)^{1/3}\hbar\bar{\omega}$, where $\bar{\omega}\equiv (\omega_x\omega_y\omega_z)^{1/3}$. With a typical total number of atoms $N\simeq 2.5\times 10^5$, $E_F\simeq k_B\times 2.0\,\mu$K.

Given the time-dependent volume integrals of $\Delta p$ and $\zeta_B$,
Eq.~\ref{eq:3.1} is  easily integrated using the initial conditions $\langle {\mathbf{r}}^2\rangle_0$ and $\partial_t\langle {\mathbf{r}}^2\rangle_0=0$. To clearly demonstrate scale-invariant expansion after the optical trap is extinguished, we integrate Eq.~\ref{eq:3.1} in two steps. First, we integrate the magnetic contribution, $\Delta\langle {\mathbf{r}}^2\rangle_{mag}$ determined from
\begin{eqnarray}
\frac{d^2}{dt^2}\frac{m\Delta\langle {\mathbf{r}}^2\rangle_{mag}}{2}&=&-2\,m\,\omega_{mag}^2
\left\{\langle y^2\rangle_0 \,[b_y^2(t)-1]+\right.\nonumber\\
& &\hspace*{-0.70in}\left.\langle z^2\rangle_0\,[ b_z^2(t)-1]-2\langle x^2\rangle_0\,[b_x^2(t)-1]\right\}.
\label{eq:3.1Mag}
\end{eqnarray}
We employ homogeneous initial conditions for the magnetic contribution, $\Delta\langle {\mathbf{r}}^2\rangle_{0\,Mag}=0$ (which does not change $\langle {\mathbf{r}}^2\rangle_0$) and $\partial_t\Delta\langle {\mathbf{r}}^2\rangle_{0\,Mag}=0$. The time dependent expansion scale factors $b_i(t)$ are  determined by fitting the measured cloud radii as a function of time after release, using the hydrodynamic equations, Eq.~\ref{eq:xsqddot2} in a scaling approximation~\cite{CaoViscosity,CaoNJP}, i.e., $\langle x_i^2\rangle=\langle x_i^2\rangle_0\,b_i^2(t)$ and $\langle v_i^2\rangle=\langle x_i^2\rangle_0\,\dot{b}_i^2(t)$. As the magnetic contribution to $\langle {\mathbf{r}}^2\rangle$ arising from Eq.~\ref{eq:3.1Mag} is only a few percent, the expansion factors are readily determined with sufficient precision. After integration, the quantity $\Delta\langle {\mathbf{r}}^2\rangle_{mag}$ is subtracted from the measured $\langle {\mathbf{r}}^2\rangle$ data to determine the effective $\langle {\mathbf{r}}^2\rangle$, which then expands according to the remaining terms in Eq.~\ref{eq:3.1},
\begin{eqnarray}
\frac{d^2}{dt^2}\frac{m\langle {\mathbf{r}}^2\rangle}{2}&=&\langle {\mathbf{r}}\cdot\nabla U_{opt}\rangle_0
+\frac{3}{N}\int d^3{\mathbf{r}}\,[\Delta p-\Delta p_{\,0}]\nonumber \\
& &-3\,\hbar\,\langle\alpha_B\,\nabla\cdot{\mathbf{v}}\rangle.
\label{eq:3.1Opt}
\end{eqnarray}
The first term is the optical trap contribution, which is dominant and leads to a $t^2$ scaling for the mean square cloud radius of a resonantly interacting gas or for a ballistic gas, with the initial condition $\langle {\mathbf{r}}^2\rangle_0$. The remaining small pressure change  and bulk viscosity terms can be integrated separately, using the scale factors $b_i(t)$ and the same homogeneous initial conditions as for the magnetic contribution. These contributions are described in  more detail below.

\subsection{High Temperature Approximation to $\Delta p$}
\label{sec:highT}

We study the regime away from the Feshbach resonance by using in Eq.~\ref{eq:3.1} a nonzero correction to the pressure $\Delta p$. We determine the energy $\widetilde{E}$ and initial cloud temperature $T_0$, using
\begin{equation}
\widetilde{E}=3\,k_B T_0=3\left\langle z\frac{\partial U_{total}}{\partial z}\right\rangle_0=\langle {\mathbf{r}}\cdot\nabla U_{total}\rangle_0,
\label{eq:temperature}
\end{equation}
which follows from force balance in the trapping potential and $p=n(0) k_B T_0$ in the high temperature limit. This approximation is adequate for evaluating $\Delta p$ in the second virial approximation, as described below. We discuss the measurement of $\langle z\partial_zU_{total}\rangle_0$ using the axial cloud profile in \S~\ref{sec:parameters}.

As a consequence of energy conservation, only the {\it difference} between the pressure at time $t$ and at time $t=0$, i.e., $\Delta p-\Delta p_{\,0}$, appears in Eq.~\ref{eq:3.1}. Hence, any {\it static} contribution to $\Delta p$ has no effect. In the high-temperature limit, we can evaluate $\Delta p$ using a virial expansion~\cite{HoMuellerHighTemp}. To determine $\Delta p$, we make the assumption that contributions to $\Delta p$ that require three-body and higher order processes to maintain equilibrium are {\it frozen} at their initial values over the time scale of the expansion, and do not contribute to $\Delta p-\Delta p_{\,0}$. In particular, the molecular contribution to the second virial coefficient~\cite{HoMuellerHighTemp} requires  three-body collisions to populate and depopulate the molecular state as the gas expands and cools in the translational degrees of freedom.  Therefore, the molecular contribution is expected to be negligible.

We evaluate $\Delta p$ in the high temperature limit to second order in the fugacity~\cite{HoMuellerHighTemp} , where $\Delta p$ is given by
\begin{equation}
\Delta p=p-\frac{2}{3}\,{\cal E}=-\frac{\sqrt{2}}{3}\,n\,k_BT\left(T\frac{\partial b_2}{\partial T}\right)(n\lambda_T^3).
\label{eq:6.3c}
\end{equation}
Here $\lambda_T\equiv h/\sqrt{2\pi mk_BT}$ is the thermal wavelength and $b_2$ is the part of the second virial coefficient that describes two-body interactions. Ignoring the molecular contribution, which is frozen on the short time of the expansion as discussed above, we take
\begin{equation}
b_2(x)=-\frac{sgn[a_S]}{2}\,e^{x^2}{\rm erfc}(x),
\label{eq:10.1}
\end{equation}
where ${\rm erfc}(x)=1-\frac{2}{\sqrt{\pi}}\int_0^x dx'\,e^{-x'^2}$ and $x=\frac{\lambda_T}{|a_S|\sqrt{2\pi}}$, with $a_S$ the s-wave scattering length.

As $\Delta p$ causes only a small perturbation to the flow, we make an adiabatic approximation for the temperature, $T=T_0\,\Gamma^{-2/3}$, where $T_0$ is the initial temperature of the trapped cloud and $\Gamma=b_xb_yb_z$ is the volume scale factor, i.e., the density $n$ of the expanding gas scales as $1/\Gamma$. We determine $\Gamma$ by fitting the aspect ratio data with a scaling approximation to the hydrodynamics~\cite{CaoViscosity,CaoNJP}, using the shear viscosity as the only free parameter. Then, $x=x_0\,\Gamma^{1/3}$, where $x_0=\frac{\lambda_{T_0}}{|a_S|\sqrt{2\pi}}$. Using the high temperature and harmonic approximation for the energy per particle, $\widetilde{E}=3\,k_B\,T_0$ and $E_F=\frac{\hbar^2 k_{FI}^2}{2m}=(3N)^{1/3}\hbar\bar{\omega}$, the Fermi energy of an ideal gas at the cloud center, we have
\begin{eqnarray}
x&=&x_0\,\Gamma^{1/3}\nonumber\\
x_0&=&\frac{\sqrt{6}}{k_{FI}|a_S|}\left(\frac{E_F}{\widetilde{E}}\right)^{1/2},
\label{eq:10.3}
\end{eqnarray}
where $k_{FI}=\sqrt{2m\,E_F/\hbar^2}$ is the Fermi wavevector of an ideal gas at the trap center.
Note that $E_F$ is measured using the total atom number and the oscillation frequencies in the trap, $\bar{\omega}\equiv(\omega_x\omega_y\omega_z)^{1/3}$, where the $\omega_i$ are given in \S~\ref{sec:parameters}.

Now, $T\frac{\partial b_2}{\partial T}=-xb'_2/2$, where $b_2'(x)\equiv sgn[a_S]\,f_2'(x)$ with
\begin{equation}
f_2'(x)\equiv\frac{1}{\sqrt{\pi}}-xe^{x^2}{\rm erfc}(x).
\label{eq:10.2}
\end{equation}
Integrating over the trap volume, and using the adiabatic approximation for the temperature and a scaling approximation for the density, we obtain
\begin{equation}
\frac{1}{N}\int d^3{\mathbf{r}}\,\Delta p=\frac{\sqrt{2}}{3}\frac{k_B T_0}{\Gamma^{2/3}}\bar{z}\,x\,b_2'(x),
\label{eq:7.4}
\end{equation}
where the trap-averaged fugacity $\bar{z}$ is an adiabatic invariant. For a gaussian density profile,
\begin{equation}
\bar{z}\equiv\frac{1}{N}\int d^3{\mathbf{r}}\,n\left(\frac{n\lambda_{T_0}^3}{2}\right)=\frac{9}{4\sqrt{2}}\left(\frac{E_F}{\widetilde{E}}\right)^3.
\label{eq:7.5}
\end{equation}
In Eq.~\ref{eq:10.3} and Eq.~\ref{eq:7.5}, we have made the harmonic approximation $\omega_x^2\langle x^2\rangle_0=\omega_y^2\langle y^2\rangle_0=\omega_z^2\langle x^2\rangle_0$ and we have used the high temperature approximation $\widetilde{E}= 3\,k_B\,T_0$.

To use Eq.~\ref{eq:3.1} to determine $\langle {\mathbf{r}}^2\rangle$ as a function of time,  the volume integral Eq.~\ref{eq:7.4} is written as,
\begin{equation}
\frac{1}{N}\int d^3{\mathbf{r}}\,\Delta p=
\frac{\widetilde{E}\sqrt{6}}{4}\left(\frac{E_F}{\widetilde{E}}\right)^{7/2}
\frac{\Gamma^{-1/3}}{k_{FI}a_S}\,f_2'(x),
\label{eq:8.2}
\end{equation}
where the time dependence of $x$ is determined by Eq.~\ref{eq:10.3} and $f_2'(x)$ is given by Eq.~\ref{eq:10.2}.
We note that the leading dependence of $\Delta p$ on the Fermi energy is $E_F^{7/2}/k_{FI}\propto E_F^3$, which is proportional to $N$ the total atom number, as it should be. We have used $sgn[a_S]/|a_S|=1/a_S$ to explicitly show that the volume integral of $\Delta p$ changes sign with the scattering length $a_S$ as the bias magnetic field is tuned across the Feshbach resonance.

As discussed above, the net pressure correction in Eq.~\ref{eq:3.1} is $\Delta p-\Delta p_{\,0}$. Hence, we also evaluate Eq.~\ref{eq:8.2}  in the limit $t=0$, where $\Gamma\rightarrow 1$ and $x\rightarrow x_0$. As $|\Delta p|\leq |\Delta p_{\,0}|$ for all $t$, the net pressure correction is positive (negative) when $\Delta p$ is negative (positive). Then, compared to the resonant case, where $1/(k_{FI}a_S)=0$, the cloud is expected to expand more rapidly when the scattering length $1/(k_{FI}a_S)<0$  and more slowly when $1/(k_{FI}a_S)>0$, as observed in the experiments (see the main text).

\subsection{Bulk Viscosity}
\label{sec:bulk}
The bulk viscosity is positive for finite $a_S$  and  must vanish in the scale-invariant regime, where $|a_S|\rightarrow \infty$. Hence, to leading order in $1/a_S$, the bulk viscosity must be {\it quadratic} in $\frac{1}{k_Fa_S}$. Using dimensional analysis, the bulk viscosity then takes the general form
\begin{equation}
\zeta_B=\hbar\,n\,\frac{f_B(\theta)}{(k_Fa_S)^2},
\label{eq:bulk1}
\end{equation}
where $\hbar\,n$ is the natural unit of viscosity, $k_F\propto n^{1/3}$ is the local Fermi wavevector and $f_B(\theta)$ is a dimensionless function of the reduced temperature $\theta\equiv T/T_F(n)$, where $k_B\,T_F(n)\equiv\epsilon_F(n)\propto n^{2/3}$ is the local Fermi energy. As discussed in the main text, by using an adiabatic approximation for $\theta$, one obtains the  trap-averaged bulk viscosity coefficient, which takes the form

\begin{equation}
\bar{\alpha}_B(t)=\bar{\alpha}_B(0)\,\Gamma^{2/3}(t),
\label{eq:bulk2}
\end{equation}
where the $\Gamma^{2/3}$ factor arises from the $1/k_F^2$ scaling.

The bulk viscosity provides the {\it only} contribution proportional to $\frac{1}{a_S^2}$, while the $\frac{1}{a_S^2}$ contribution to the volume integral of  $\Delta p-\Delta p_{\,0}$ generally vanishes, as we now show. Using dimensional analysis, the most general $1/(k_Fa_S)^2$ contribution to $\Delta p$, which we define as $\Delta p_2$, must be of the form
$$\Delta p_2=n\,\epsilon_F(n)\,\frac{f_p(\theta)}{(k_Fa_S)^2},$$

\noindent where $\epsilon_F(n)$ is the local Fermi energy and $f_p(\theta)$ is a dimensionless function of the reduced temperature. As $\frac{\epsilon_F(n)}{(k_Fa_S)^2}=\frac{\hbar^2}{2ma_S^2}$, $\frac{1}{N}\int d^3{\mathbf{r}}\,\Delta p=\frac{\hbar^2}{2ma_S^2}\langle f_p(\theta)\rangle$ is time-independent, since the number of atoms in a volume element is conserved  during the expansion and $\frac{1}{N}\int d^3{\mathbf{r}}\,n\,f_p(\theta)$ is constant in the adiabatic approximation. Hence, the $\frac{1}{a_S^2}$ part of $\Delta p-\Delta p_{\,0}$ in Eq.~\ref{eq:3.1} vanishes and the bulk viscosity provides the only time-dependent $\frac{1}{a_S^2}$ contribution, which increases as $\Gamma^{2/3}$, according to Eq.~\ref{eq:bulk2}.

The form of $\bar{\alpha}_B(0)$ in Eq.~\ref{eq:bulk2} is obtained from Ref.~\cite{DuslingSchaferBulk}, where bulk viscosity is predicted  in the high temperature limit. To second order in the fugacity $z$,
\begin{equation}
\zeta_B=\widetilde{c}_B\,\left(\frac{\lambda_T}{a_S}\right)^2\,\frac{\hbar}{\lambda_T^3}\,z^2,
\label{eq:bulkpredict}
\end{equation}
where we can approximate $z=n\,\lambda_T^3/2$. Dusling and Sch\"{a}fer~\cite{DuslingSchaferBulk} give $\widetilde{c}_B=\frac{1}{24\pi\sqrt{2}}$.

Integrating over the trap volume, we obtain $\bar{\alpha}_B(t)=\bar{\alpha}_B(0)\,\Gamma^{2/3}(t)$ as in Eq.~\ref{eq:bulk2} with
\begin{equation}
\bar{\alpha}_B(0)=\widetilde{c}_B\,\frac{27\pi}{2\sqrt{2}}\frac{1}{(k_{FI}a_S)^2}
\left(\frac{E_F}{\widetilde{E}}\right)^4\equiv c_B\,\left(\frac{E_F}{\widetilde{E}}\right)^4.
\label{eq:bulkviscoef}
\end{equation}
Here, the value of $c_B$ based on the high-temperature prediction is
\begin{equation}
c_B=\frac{9}{32}\frac{1}{(k_{FI}a_S)^2}.
\label{eq:bulkpred}
\end{equation}

\subsection{Measuring $\Delta p$ and the Bulk Viscosity.}
\label{sec:parameters}
 We measure the effect of $\Delta p$ on the flow at finite scattering length and estimate the bulk viscosity  by tuning the bias magnetic field $B$ away from resonance and observing the departure of $\langle{\mathbf{r}}^2\rangle$ from ballistic flow. For ballistic flow,
\begin{equation}
\langle{\mathbf{r}}^2\rangle=\langle{\mathbf{r}}^2\rangle_0+\frac{t^2}{m}\,\langle{\mathbf{r}}\cdot\nabla U_{opt}\rangle_0.
\label{eq:scaleinv2}
\end{equation}
The $t^2$ form of Eq.~\ref{eq:fitc1c0} is exactly valid for a resonantly interacting gas and for a ballistic (noninteracting) gas. However, for finite scattering length at an arbitrary bias magnetic field $B$, Eq.~\ref{eq:3.1Opt} shows that $\langle {\mathbf{r}}^2\rangle$ does not expand as $t^2$. However, for the range of interaction strengths studied in our experiments,  $\Delta p$ and the bulk viscosity produce only small perturbations to the flow. For this reason, we can continue to parameterize the time evolution of $\langle{\mathbf{r}}^2\rangle$ using
\begin{equation}
\langle {\mathbf{r}}^2\rangle=c_0+c_1\,t^2.
\label{eq:fitc1c0}
\end{equation}
We determine $c_1=c_1[B]$ in Eq.~\ref{eq:fitc1c0} from the fit to the data and compare the ratio $c_1[B]/c_0$ for finite $1/a_S$ to that obtained at resonance. This method avoids utilizing model-dependent expansion factors for the cloud radii in the data analysis.

All of the measured $\langle {\mathbf{r}}^2\rangle$ are corrected for the effective potential arising from magnetic field curvature  by subtracting the  magnetic field contribution, Eq.~\ref{eq:3.1Mag}. Then we determine the {\it effective} ratio $(c_1[B]/c_0)/(c_1[834]/c_0)$, described in more detail below and shown in Fig.~\ref{fig:c1c0}. If $\Delta p$ and $\zeta_B$ were zero at all magnetic fields, then this ratio would be unity everywhere, corresponding to the  black line in the figure. The systematic deviation from unity arises from finite $\Delta p$ and $\zeta_B$, where the red dots (top) show data at 986 G where $a_S<0$ and the blue dots (bottom) show data for 760 G, where $a_S>0$. In the following sections, we describe the analysis in more detail.

\subsubsection{Energy Scale}

We begin by noting that the ratios $c_1/c_0$ are energy dependent. To provide an energy scale for all of the experiments, we use
\begin{equation}
\widetilde{E}\equiv\langle{\mathbf{r}}\cdot\nabla U_{total}\rangle_0.
\label{eq:energy}
\end{equation}
$\widetilde{E}$ is twice the  potential energy per particle in the harmonic oscillator approximation. For the resonantly interacting gas, where the virial theorem holds, $\widetilde{E}$ is precisely the energy of the cloud for a harmonic trapping potential. For an anharmonic trap, the virial theorem gives the total energy for the resonantly interacting gas in terms of the trapping potential~\cite{ThermoLuo}, but  at finite scattering length, the relation between the total energy and the trapping potential energy is scattering length dependent~\cite{WernerVirial}. By using  $\widetilde{E}$ to characterize the energy, we avoid the scattering length dependence.
As discussed in \S~\ref{sec:highT}, for evaluating $\Delta p$ and $\zeta_B$ in the high temperature limit, we determine the initial temperature of the cloud $T_0$  from $\widetilde{E}=3\,k_B\,T_0$.

\subsubsection{Determining $\langle{ \mathbf{r}}\cdot\nabla U_{opt}({\mathbf{r}})\rangle_0$}
\label{sec:xdotgradU}

For precise measurements, it is necessary to determine both the harmonic value and the anharmonic corrections to $\langle {\mathbf{r}}\cdot\nabla U_{opt}({\mathbf{r}})\rangle_0$.
We recall that the total trapping potential takes the form
\begin{equation}
U_{total}({\mathbf{r}})= U_{opt}({\mathbf{r}})+U_{mag}({\mathbf{r}}),
\label{eq:totalU}
\end{equation}
where $U_{opt}$ arises from the optical trap and $U_{mag}$ arises from the magnetic field curvature, Eq.~\ref{eq:Mag}.

 We note that for a scalar pressure $p$, force balance for the trapped cloud requires  $\int d^3{\mathbf{r}}\, p=\langle x\partial_x U_{total}\rangle_0 =\langle y\partial_y U_{total}\rangle_0=\langle z\partial_z U_{total}\rangle_0$. Then,
\begin{equation}
\langle {\mathbf{r}}\cdot\nabla U_{total}({\mathbf{r}})\rangle_0=3\,\left\langle z\frac{\partial U_{total}}{\partial z}\right\rangle_0
\end{equation}
and
\begin{equation}
\langle {\mathbf{r}}\cdot\nabla U_{opt}({\mathbf{r}})\rangle_0=3\,\left\langle z\frac{\partial U_{total}}{\partial z}\right\rangle_0 -\langle {\mathbf{r}}\cdot\nabla U_{mag}({\mathbf{r}})\rangle_0 .
\label{eq:xdotgradU}
\end{equation}
Since $\langle x^2\rangle_0$ and $\langle y^2\rangle_0$ are small compared to $\langle z^2\rangle_0$, the last term is just $m\,\omega_{mag}^2\,\langle z^2\rangle_0$. Using Eq.~\ref{eq:totalU} in Eq.~\ref{eq:xdotgradU}, we then have
\begin{equation}
\langle {\mathbf{r}}\cdot\nabla U_{opt}({\mathbf{r}})\rangle_0=3\,\left\langle z\frac{\partial U_{opt}}{\partial z}\right\rangle_0 +2\,m\,\omega_{mag}^2\,\langle z^2\rangle_0  .
\label{eq:xdotgradUopt}
\end{equation}

The harmonic oscillator frequency for atoms in the optical trapping potential is generally energy dependent, because the optical trap is less confining at higher energy, causing the frequency to decrease. We model this by writing
\begin{equation}
\left\langle z\frac{\partial U_{opt}}{\partial z}\right\rangle_0=m\,\omega^2_{z\,opt}\,\langle z^2\rangle_0\,h_A(\langle z^2\rangle_0),
\end{equation}
where $m\,\omega^2_{z\,opt}\,\langle z^2\rangle_0$ arises from the harmonic trapping potential. Here $h_A$ is the anharmonic correction factor,
\begin{equation}
h_A(\langle z^2\rangle_0)\equiv 1-\lambda_1\,\langle z^2\rangle_0,
\label{eq:hA834}
\end{equation}

Hence,
\begin{eqnarray}
\frac{\langle {\mathbf{r}}\cdot\nabla U_{opt}({\mathbf{r}})\rangle_0}{m}&=&
3\,\omega^2_{z\,opt}\,\langle z^2\rangle_0\,h_A(\langle z^2\rangle_0)\nonumber\\
& &+2\,\omega_{mag}^2\,\langle z^2\rangle_0  .
\label{eq:xdotgradUopt2}
\end{eqnarray}

\subsubsection{Determining the Anharmonic Correction}
\label{sec:anharmonic}

We use Eq.~\ref{eq:xdotgradUopt2} to measure the anharmonic correction factor  $h_A=1-\lambda_1\langle z^2\rangle_0$, by making measurements of $\langle {\mathbf{r}}^2\rangle=c_0+c_1[B]\,t^2$ for  both an ideal gas at $B=528$ G and a resonantly interacting gas at $B=834$ G for several different energies. For these two cases, $c_1=\langle {\mathbf{r}}\cdot\nabla U_{opt}({\mathbf{r}})\rangle_0/m$ and  $c_0$ determines $\langle z^2\rangle_0$ from
\begin{equation}
c_0=\langle {\mathbf{r}}^2\rangle_0=\langle z^2\rangle_0\,\left(1+\frac{\omega_z^2}{\omega_x^2}
+\frac{\omega_z^2}{\omega_y^2}\right),
\label{eq:c0}
\end{equation}
where $\omega_z^2\equiv \omega_{z\,opt}^2+\omega_{mag}^2$ and the quantity in parentheses is  close to unity in the our experiments.
Solving Eq.~\ref{eq:xdotgradUopt2} for $h_A$, we obtain
\begin{equation}
h_A[B,\langle z^2\rangle_0]=\frac{c_1[B]}{3\,\omega^2_{z\,opt}\langle z^2\rangle_0}-\frac{2\,\omega_{mag}^2[B]}{3\,\omega^2_{z\,opt}}.
\label{eq:anharmonicfit}
\end{equation}

 For a resonantly interacting cloud or an ideal gas in a harmonic optical trap, we would have $h_A[834,\langle z^2\rangle_0]=1$, by construction. Instead we find that $h_A[834,\langle z^2\rangle_0]$  decreases  with increasing $\langle z^2\rangle_0$, i.e., $\lambda_1>0$, as expected for a correction arising from trap anharmonicity, where the quartic terms in the optical trapping potential decrease the average oscillation frequency.

The optical frequency $\omega_{z\,opt}$ in Eq.~\ref{eq:anharmonicfit} is most precisely determined at 834 G by demanding $h_A=1$ for energies close to the  ground state, where the anharmonic correction is small and the resonantly interacting gas is nearly a pure superfluid.

\begin{figure}
\begin{center}\
\includegraphics[width=3.0in]{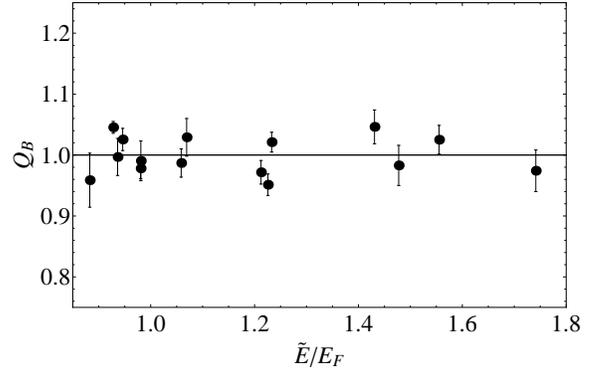}
\caption{Resonantly interacting Fermi gas. The black dots are obtained from the fit to individual expansion curves using $\langle {\mathbf{r}}^2\rangle=c_0+c_1\,t^2$ to determine $Q_B$, Eq.~\ref{eq:ratio}. The black horizontal line denotes the ideal value of unity. \label{fig:c1c0Resonance}}
\end{center}
\end{figure}

The slope $\lambda_1$  is determined by measuring the ballistic expansion of the  noninteracting Fermi gas at 528 G as a function of initial cloud size, using the same method as for the resonantly interacting gas. In the experiments, we find that the $\lambda_1$ obtained from the $h_A$ data for ballistic expansion at 528 G is within 10\% of that obtained from the $h_A$ data for the highest energies of the resonantly interacting gas. By construction, the quantity $h_A[834,\langle z^2\rangle_0]/(1-\lambda_1\langle z^2\rangle_0)$ is then unity at all energies $\widetilde{E}$, corresponding to the black horizontal line in Fig.~\ref{fig:c1c0Resonance}.

\subsubsection{Results}

For the resonantly interacting gas at all initial energies, we use the linear anharmonic correction $h_A$ and $\omega_{z\,opt}$, determined as described above, to predict $\langle {\mathbf{r}}\cdot\nabla U_{opt}({\mathbf{r}})\rangle_0$ according to Eq.~\ref{eq:xdotgradUopt2}. Using this in Eq.~\ref{eq:3.1Opt} and $\Delta p=0$, we fit the expansion data for each energy $\widetilde{E}$ and find that the corresponding bulk viscosity is very small. The energy-averaged bulk viscosity coefficient is consistent with zero, as described in the main text.
\begin{figure}
\begin{center}
\includegraphics[width=3.0in]{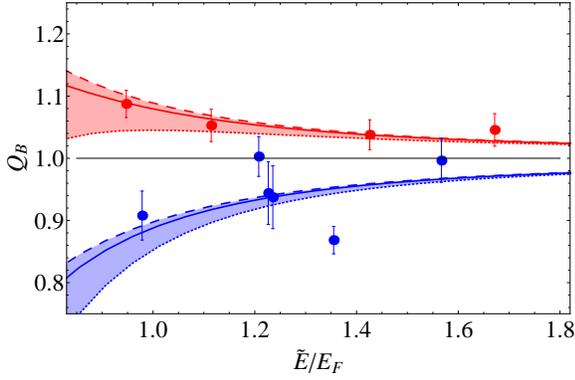}
\end{center}
\caption{Pressure change $\Delta p$ and bulk viscosity $\zeta_B$ contributions to conformal symmetry breaking as a function of energy.  The data is fit with $\langle {\mathbf{r}}^2\rangle=c_0+c_1\,t^2$. The effective ratio $(c_1/c_0)/(c_1/c_0)_{834}$ (given in detail by Eq.~\ref{eq:ratio}) is shown for the resonantly interacting gas $1/(k_{FI}a_S)=0$ (black line-theory),  for $1/(k_{FI}a_S)=-0.59$ (top, red dots) and  for $1/(k_{FI}a_S)=+0.61$ (bottom, blue dots).   Solid curves top and bottom show the best fit, where $\lambda_p=1.07(0.25)$ and $\lambda_B=0.20(0.55)$, see Fig.~\ref{fig:Contour}. The dashed (dotted) curves show the predictions for $\lambda_p=1.06$ and $\lambda_B=0$ ($\lambda_B=1$), to illustrate the effect of the bulk viscosity.  \label{fig:c1c0}}
\end{figure}

For the off-resonant studies at a bias field $B$, we again fit  $\langle {\mathbf{r}}^2\rangle=c_0+c_1\,t^2$ to the expansion data. We can still determine $h_A[B,\langle z^2\rangle_0]$ from Eq.~\ref{eq:anharmonicfit}. However, since $c_1[B]$ is modified by the nonzero $\Delta p$ and $\zeta_B$, $h_A[B,\langle z^2\rangle_0]$  deviates from $1-\lambda_1\langle z^2\rangle_0$. Therefore, we characterize the flow using the ratio,
\begin{equation}
Q_B\equiv \frac{h_A[B,\langle z^2\rangle_0]}{1-\lambda_1\langle z^2\rangle_0}
\label{eq:ratio}
\end{equation}
for each energy $\widetilde{E}$. By construction, this ratio is unity for $B=834$ G, corresponding to the black horizontal line in Fig.\ref{fig:c1c0}. For $B\neq 834$ G, where $\Delta p\neq 0$ and $\zeta_B\neq 0$, the ratio deviates systematically from unity.

\begin{figure}
\begin{center}
\vspace*{0.25in}\includegraphics[width=3.0in]{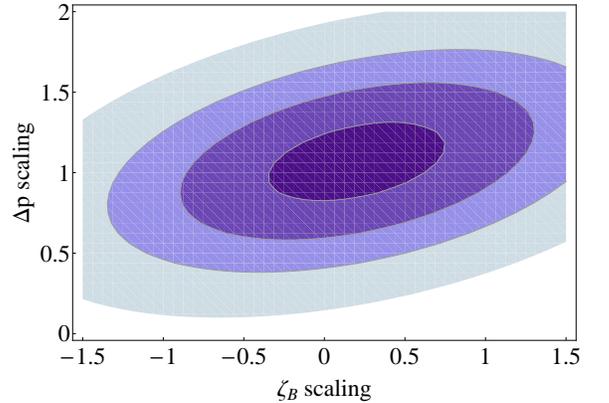}
\end{center}
\caption{Contour plot of  $\chi^2$ for all of the off-resonance data as a function of $\lambda_B$  and $\lambda_p$. The data shown in Fig.~\ref{fig:c1c0} are compared to the high temperature predictions of \S~\ref{sec:highT} and \S~\ref{sec:bulk} using two scaling parameters,  $\lambda_p$ for $\Delta p$ and $\lambda_B$ for the bulk viscosity.  \label{fig:Contour}}
\end{figure}

Fig.~\ref{fig:c1c0} shows the ratio $Q[B]$ of Eq.~\ref{eq:ratio} as a function of the energy $\widetilde{E}$. For comparison,  we use Eq.~\ref{eq:3.1Opt} to predict the corresponding ratios  for each energy as a function of two scaling parameters, $\lambda_p$ for $\Delta p$ calculated in the high temperature limit and $\lambda_B$ for the predicted high temperature bulk viscosity. The  solid lines (top and bottom)  Fig.~\ref{fig:c1c0} show the best fit to the data, see the contour plot, Fig.~\ref{fig:Contour}, where $\lambda_p=1.07(0.25)$ and $\lambda_B=0.20(0.55)$. The contribution of the bulk viscosity  appears smaller than predicted. To show the relative scale of the bulk viscosity and the $\Delta p$ corrections, the predictions for $\lambda_p=1.07$ and $\lambda_B=0$ are shown as dashed curves, while the dotted curves show the predictions for $\lambda_p=1.07$ and $\lambda_B=1$. As our $\Delta p$ model adequately describes the data, it appears that the pressure in the expanding cloud is not far from local equilibrium in the translational degrees of freedom, and  the observed breaking of scale invariance  is  primarily due the direct change in the pressure $\Delta p=p-\frac{2}{3}{\cal E}\neq 0$.


\begin{thebibliography}{23}
\bibitem[{Not()}]{NoteMag}
\bibinfo{note}{A small correction $\Delta\langle{\mathbf{r}}^2\rangle_{Mag}$
  arising from the curvature in the bias magnetic field is subtracted from the
  $\langle{\mathbf{r}}^2\rangle$ data to obtain the $\tau^2(t)$ data that is
  shown in the figure~\cite{SupportOnline}.}
\end{thebibliography}
\end{document}